# Areas of Strategic Visibility: Disability Bias in Biometrics

Position Statement, Submitted to EFC, January 2021.

Signatories[1]: Jennifer Mankoff, representing the [Center for Research and Education on Accessible Technology and Experiences](U. Washington); Devva Kasnitz, Disability Studies (City University of New York); L Jean Camp (Indiana U.); Jonathan Lazar (U. of Maryland, HCIL, Trace Center); Harry Hochheiser (U. of Pittsburgh)

**Overview.** This response to the RFI considers the potential for biometrics to help or harm disabled people[2]. Biometrics are already integrated into many aspects of daily life, from airport travel to mobile phone use. Yet many of these systems are not accessible to people who experience different kinds of disability exclusion . Different personal characteristics may impact any or all of the physical (DNA, fingerprints, face or retina) and behavioral (gesture, gait, voice) characteristics listed in the RFI as examples of biometric signals.

We define disability here in terms of the discriminatory and often systemic problems with available infrastructure's ability to meet the needs of all people [UN 2017, Oliver, 2013]. Using this definition, *"[biometrics] could either mitigate or amplify disability depending on how they are designed."* (Guo, 2019). As Whittaker and colleagues (2019) state, this is not simply a matter of algorithmic accuracy: *"…discrimination against people of color, women, and other historically marginalized groups has often been justified by representing these groups as disabled…. Thus disability is entwined with, and serves to justify, practices of marginalization."* (p. 11). It is critical that we look beyond inclusion to full and fully accommodated participation. Just being in the room is not enough, or just, when power, understanding and action are in the hands of computer scientists, business people and the many other stakeholders implementing biometric systems . This report adopts the philosophy of a recent report from the AI Now institute authored by multiple disabled disability scholars (*ibid.*), which asks *"How do we move from "inclusion" to 'agency and control,' given the increasingly proprietary nature of the technologies being created, and the centralization inherent in the current form of AI?"* Further, as Bennett and Keyes (2020) argue, we must look beyond **fairness**, which can only *"reproduce the discrimination it seeks to remedy,"* to disability **justice**, a term used by activist scholars steeped in the Black Lives Matter movement (Wong, 2020). Below we address each of the six categories discussed in the RFI, noting risks and opportunities. Ultimately, the concerns raised in this report can only be fully addressed with the addition of structural changes that will require regulation and control to exist within the for-profit system. Without the right guardrails, **both**

---

[1] The authors of this response include people with personal disability experience and people with experience in biometric and AI technology creation. This response also reflects the words of thought leaders in the field as represented in numerous cited existing reports and commentary.

[2] The language for describing people and disability is as diverse as the experience of disability itself. We reject the oversimplified binary of "identity-first" and "person-first" language. We use both, and other alternatives, based on context and goals, reflecting and respecting where modern dialogue on these issues stands.

**primary categories of use raised in the RFI (recognition and inference) can actively harm or exclude** disabled people.

## 1. Uses of Biometric Information for Recognition and Inference

The RFI points out two primary categories of use: *Recognition* and *Inference*. The benefits of such technologies are similar for people with and without impairments, however *access to such technologies is important for equitable use*. Thus, **it is critical that basic accessibility barriers with biometric technologies are addressed** (Guo, 2019). Even, if a biometric algorithm is unbiased, the interface to that algorithm, its configuration (Kane 2020), or the explanation of how it works (Wolf, 2019), may all be inaccessible. Ableist assumptions built into an application can make it inaccessible even if it meets legal standards. An example from a recent survey of disabled users of biometric systems is system timeouts that *"do not account for the slower movement speeds of people with physical disabilities"* (*ibid.*), including doors closing too fast or locking before a person can get to them; as well as timeouts in voice menus; bathroom lights; and vending machines/ATMs. The simple inability to hold still enough for biometrics to register is also often overlooked. Similarly, Kane (2020) describes how systems may not address differences in height (*e.g.,* for wheelchair users) or other physical factors such as strength, stamina, or range of motion. We note that while "Universal Design" of biometric systems is not feasible, just as one toilet height serves "most" but not all wheelchair riders, the *flexibility to adapt* can ensure accessibility to all. Such an approach avoids creating separate systems for a subset of people, which risks inequitable outcomes over time as one system is updated and others are not (Lazar, 2015).

A second and related concern in any data-driven biometric system is the fact that **data sets used to train biometric systems are biased:** they rarely if ever have a comprehensive representation of the range of people they might encounter. A person might have unusual or missing limbs and not have a fingerprint, or walk differently, or speak differently than the system expects, and thus be unable to access services tied to recognition of fingerprints, gait, or voice. Further, the oversimplification of disability experience into single diagnoses or symptoms often excludes people with multiple impairments.

## 2. Procedures for and results of data-driven and scientific validation of biometric technologies

The RFI asks for *"procedures data-driven and scientific validation of biometric technologies."* It is critical for such procedures to reflect the full range of persons who may be impacted by such technologies. *Measures*, and their *validity*, are important, but *what is measured* is equally important. Bias in biometric data, such as overlooking multiple co-occurring impairments, or using a voice print for identification when voices are missing, machine-produced, or unusual enough to not be recognized as human, can take several forms that all must be considered if we are to rectify the errors that result.

- **Human or Machine Bias:** Biased human perceptions of disability can accidentally be captured during data entry (Trewin, 2018). For example, suppose crowd workers are asked to label affect in images of disabled people without proper training. Similarly, outlier detection represents a form of machine bias. Algorithms that flag or remove outliers, either at training time or at inference/detection time, may erroneously exclude people who are under-represented in the data (Guo, 2019).
- **Unrepresentative and Overly Simplified Data** When groups are historically marginalized and underrepresented, this is *"imprinted in the data that shapes AI systems… Those who have borne discrimination in the past are most at risk of harm from biased and exclusionary AI in the present."* (Whittaker, 2019, p. 8). **Addressing bias is not a simple task of increasing the number of categories represented** *(ibid.)*. Impairment is not static, homogenous, nor do people only have one impairment. One person may have many impairments with synergistic effects. For example, facial recognition is less successful for older adults with dementia (Taati, 2019) and gait recognition accuracy differs based on age and gender (the study did not include disability). Yet, older adults are significantly under-represented in AI data sets (Park, 2021b). Other intersecting non-disability characteristics, such as accented speech, or technology fluency, can further impact data (Whittaker, 2019; Trewin, 2018). Additionally, the same impairment may vary across individuals or change based on age of onset, or over time (Kasnitz, 2012). It is critical to collect data about people from multiple contexts with multiple impairments over multiple timescales, rather than assuming a single fixed experience of disability.
- **Measurement Error** Measurement error can further exacerbate bias (Trewin, 2018). For example, a Fitbit may not recognize wheelchair activity as exercise, a bias in its measure of activity. Guo *et al*. (2019) provide an extensive list of examples for each major class of biometric sensing. Guo *et al. (*2019*)* discuss how facial mobility, emotion expression, and facial structure impact *detection*, *identification*, *verification*, and *analysis* (*e.g.,* emotion analysis)); how body motion and shape impact "body recognition" (*e.g.,* activity detection); handwriting analysis; and speaker analysis.

Addressing bias in biometric data requires assessment methods that can uncover bias. Aggregate metrics can hide performance problems in under-represented groups (Besmira, 2018). Many algorithms maximize these metrics and thus not only fail to recognize bias, but also to address it (Guo, 2019). For example, algorithms that eliminate, or reduce the influence, of outliers are more likely to eliminate disabled people because of the heterogeneity of disability data. Trewin (2018) covers several alternative options for assessment, and highlights *"individual fairness,"* defined as comparing performance (outcomes) between people who are similar, where similarity is defined using metrics that are chosen not to encode bias. For example, movement speed might favor a wheelchair user and exercise variety might favor people who do not have chronic illness; while measures of exertion time might be a similarity metric that covers a wide variety of different types of people. Defining such unbiased metrics requires careful thought and domain knowledge, and scientific research will be essential to defining appropriate procedures for this.

## 3. Security considerations in making biometric technologies accessible

Biometric systems used by people with disability have all of the same risks as anyone faces regarding data breaches and other aspects of privacy and security [Ritter, 2021]. However, ableism and other biases embedded in society raise additional disability-specific risks.

**Privacy and Security.** The risk to disabled people of data disclosure can include direct harms such as denial of insurance and medical care, or threaten employment (Whittaker, 2019, p. 21). Any system that can detect disability can also track its progression over time, possibly disclosing disability even before a person knows themselves that they have a diagnosis (or incorrectly labeling someone). Yet this is an uneven flow of information -- the person being labeled may not even know it is happening, or even if they do it may not be voluntary, as suggested by Whittaker et al (*ibid.*).  Further, small sample sizes for people with rare disabilities may make data security more difficult. For example, an algorithm may learn to recognize the disability, rather than the individual, reducing security when used for access control, allowing multiple people with similar impairments to access the same data.

**Diagnosing, or pathologizing disability or illness.** From DNA to voice to gesture and gait, the data biometric systems collect can easily be used to learn about disability. This is not just theory -- for example, Whittaker et al (2019) document how HireVue, an AI based video interviewing company has a patent on file to detect disability (Larsen, 2018), despite the fact that Title I of the Americans with Disabilities Act (ADA) forbids asking about disability status in a hiring process (42 U.S.C. § 12112(a)) and also forbids *"using qualification standards, employment tests or other selection criteria that screen out or tend to screen out an individual with a disability"* (42 U.S.C. § 12112(b)(6)). HireVue's intent is to reduce algorithmic discrimination, however, such information could easily be used, without consent, to deny access to housing, jobs, or education. Disability identification is spreading, including detecting Parkinsons from gait (Das, 2012), and mouse movement (Youngmann, 2019), and detecting autism from home videos (Leblanc, 2020). While disability detection may have value, **the potential for abuse of these tools makes regulation a necessity.**

Further, as Whittaker et al (2019, p. 21) point out, algorithms often define disability entirely in historical medical terms, potentially replicating biases (Bennett, 2020), that then cause a person to go unrecognized and thus to be gatekept out of support systems.  This is inconsistent with U.S. Federal law, since the ADA does not require a diagnosis for disability protections, simply that a person be regarded as having a disability ( 42 U.S.C. § 12101 (a)(1)). The underlying idea is brilliantly progressive, albeit often under attack: **Legally, if you are treated as disabled, you are disabled. Yet biometrics cannot detect how people are treated. Biometrics must never be considered sufficient, nor required as mandatory, for disability identification or service eligibility,** but it will be proposed for both in systems seeking easy answers to complex phenomena.

## 4. Exhibited and potential harms of a particular biometric technology

Even if accessibility concerns with interfaces to biometrics are addressed, there are numerous additional disability-related risks, including incorrect recognition of faces, fingerprints, and speech; and incorrect inferences about activity and gender (Kane, 2020), raising several severe areas of concern and potential risk which have been laid out in detail in the literature (*e.g.,* Whittaker, 2019).

**Defining, or enforcing "normality" based on a biased data set**. As Whittaker (2019) argue, norms are baked deeply into algorithms which are designed to learn about the most common cases. As human judgment is increasingly replaced by biometrics, "norms" become more strictly enforced. There will always be outliers, these outliers will face higher error rates, and they will disproportionately represent and misrepresent people with disability. Resulting errors can impact **allocation of a resource** (Guo, 2019). Biometrics already are being used to track the use and allocation of assistive technologies, from CPAP machines for people with sleep apnea (Araujo 2018) to prosthetic legs (as described by [Jullian Wiese in Granta](#) and uncovered in Whittaker et al 2019), deciding who is "compliant enough" to deserve them. Recent changes in California's automating billing procedures for In Home Supported Services require navigating inaccessible phone or online AI verification procedures, further impacting resource access.

**Defining, or enforcing what it means to be "human"**. From government services to education, healthcare, finances (including ATM use) and even basic computer security, access to services today often depends on passing biometric tests. Yet, many biometric systems gatekeep access based on either individual identity, identity as a human, or class of human, such as "old enough to buy cigarettes." When biometric systems are not accessible, they are essentially defining a disabled person as non-human, or not enough of something with respect to the service being denied. Kane (2020) give examples, such as a participant having to falsify data because *"some apps [don't allow] my height/weight combo for my age."* Often, the only solution is to accept reduced digital security, such as the person who must ask a stranger to 'forge' a signature at the grocery store *".. because I can't reach [the tablet]"* (*ibid.*). This is not only inaccessible, it is illegal: kiosks and other technologies such as point-of-sale terminals used in public accommodations are covered under Title III of the ADA, as clearly stated by the U.S. Department of Justice (2014). At work, activity tracking may define "success" in terms that exclude disabled workers. Further, technology may simply fail to recognize that a disabled person is even present (Kane, 2020), a phenomenon they term *invisibility*, because it others and erases people. Such systems amplify existing biases internal to and across othering societal categories (Guo, 2019), reflecting and even enforcing normative categories, thus *"demarcating what it means to be a legible human and whose bodies, actions, and lives fall outside... [and] remapping and calcifying the boundaries of inclusion and marginalization (*Whittaker, 2019). The calcification of such decisions in code risks harm not only in each decision but also through **obscuring the processes for improvement** of such problematic decision making.

**Exacerbating or Causing Disability.** Whittaker et al (2019) raise concerns about how activity tracking systems may push workers to limits that increase the likelihood of work-related disability, by forcing workers to work at maximal efficiency. Even where accommodations are provided they may have unrecognized time or contextual limitations. Further, biometrics may limit access to critical care resources such as human assistance, resulting in increased risk of hospitalization or institutionalization (Lecher, 2018). These harms are exacerbated when biometric systems, by **removing the human, remove the humane nature of decision making and replace open systems with closed systems***.* Such closed systems remove control over the reasons behind decisions and obscure concerns such as whether data is representative or algorithms are erroneous or fair.

## 5. Exhibited and potential benefits of biometric technology

There are also some disability-specific benefits of biometric technology. For example, biometric technologies can provide opportunities for improved access by replacing a less accessible option. An example is that face recognition may be an easier way to handle phone security than passcode entry for someone who lacks physical dexterity. However, many of the potential benefits of biometrics for disabled domains are dependent upon input from the communities being served. Overlooking disabled peoples' expertise in their own needs risks creating systems that exacerbate harm rather than improving lives.

**Behavioral training/support for independence.** For example, biometrics have been used in commercial products to recognize affect, gaze, and other behaviors in support of autistic individuals. While marketed for their therapeutic and other benefits, the result can be highly problematic and contribute to contentious, debated practices, rather than contributing to the agency and independence of the target audience [Demo, 2017]. The stakeholders targeted, underlying beliefs guiding the app design and marketing of these apps are all sources of potential harm.

**Public safety.** People with visible disability can easily be misunderstood and even targeted by both criminals and law enforcement (Trewin, 2019). Biometrics could help to classify behaviors, or re-interpret facial cues, as non-threatening (*ibid.*). However this must be weighed against the potential of increased risk of *misinterpretation* with biased data for training, and the overall risk to society of using biometrics systems for public safety (*ibid.*)

**Diagnosis.** The potential for biometric technologies to flag a situation that may require medical intervention is well established (Trewin, 2019). However, as stated earlier in the discussion of privacy and security, this brings severe risks as well. For example, Bennett and Keys (2020) provide a case study of a system that uses biometric information to "diagnose" autism, highlighting a number of risks that a naive approach to fairness, which simply examines "*the immediate algorithmic inputs and outputs of the computer vision system,*" cannot rectify. They describe how gender bias in diagnostic methods may be replicated in a diagnostic tool; how diagnosis reinforces the medicalization of the autism; the removal of power from patients; and the lack of consideration of potential harms of diagnosis including financial cost, murder, and social consequences. They conclude by stating: *"we need a model that considers holistic, societal implications, and the way that technologies alter the life chances of those they are used by or on."*

These examples demonstrate the potential for biometrics to contribute positively to the lives of people with disabilities. However this possibility can only be realized through careful application of appropriate and inclusive design methods.

## 6. Governance programs, practices or procedures

As eloquently stated by Bennett and Keys (2019), rectifying bias through fairness is necessarily an incomplete solution. Fairness cannot rectify structural differences with its reliance on well defined traits and focus on individual identities and goals rather than holistic improvements. Instead, fair biometric systems require a nuanced understanding of issues surrounding disability justice and the lived experience of disability (Wong, 2020). Further, strong ethical standards have a profound effect on professional work, as evidenced when comparing medicine to fields like AI (Mittlestadt, 2019). Such standards go beyond policy, and developing them must be a priority going forward.

### a. Stakeholder Engagement Practices: Changing who builds biometric systems

Appropriate expertise, meaning direct input from the inception to evaluation of a project from the disability community (with appropriate compensation), is critical to successfully addressing bias without introducing new risks and errors into biometric systems. There are *"significant power asymmetries between those with the resources to design and deploy AI systems, and those who are classified, ranked, and assessed by these systems"* (Whittaker, et al, 2019, p. 9). This can improve--but only if we take steps to **ensure that disabled people are included in the design of biometric systems.** Participatory design is a critical way to include people with personal disability expertise (Quintero, 2020). However, true equity will require that people with disabilities can enter the technology workforce so that they can directly build and innovate such systems. This requires access to higher education programs; access to conferences and events where research and products are discussed, presented and shared; and accessible toolkits and development environments including for user interface development, data analysis, and general programming.

In addition, the disability community needs to form a broad coalition and organize itself to impact regulation of biometric systems; prioritization regarding where biometrics would add value; and decisions about data collection. Community representation can not only improve the range and quality of participation in data collection, but may guide the design of data collection systems and prioritization of what data to collect.

### b. Best practices for pilots or trials to inform further policy developments

As a general policy rule, algorithms that put a subset of the population at risk should not be deployed. This requires both regulatory intervention and research, at the algorithmic level (*e.g.,* developing better algorithms for handling outliers) and the application level (*e.g.*, studying the risks of harm applications might create for disabled people). Both studies and regulation must take an holistic approach that, rather than being exclusively about technology *"accounts for the context in which such technology is produced and situated, the politics of classification, and the ways in which fluid identities are (mis)reflected and calcified through such technology"* (Whittaker, 2019, p. 11). Regulatory decisions must be informed by analyses that consider all of these factors, to strongly guide industry practice.

Further, accessibility solutions must be directly implemented in existing products: It is well established in the literature that "separate but equal" technological solutions are not equitable, because there is no economic incentive to ensure equality is maintained over time (Lazar, 2015). Accessible options must also be complete and easy to use. Further, both research and regulation must look at biometrics in combinations with each other and with the non-biometric systems they are designed to replace to assess what constitutes an equitable, accessible system. Just as accessible ramps or elevators that are hidden or far away are not considered acceptable for accessibility in physical spaces, truly accessible biometric systems must not create undue burdens in digital spaces nor segregate disabled users. While a single interface may not be accessible to all people, a single, flexible *system of solutions* with appropriate accessibility support can be.

### c. Practices regarding data collection, review, management, storage and monitoring

Park *et al*. (2021a) lay out design guidelines for data collection including how to motivate participants and appropriate pay; what to communicate at data collection time, and how to make sure that data collection infrastructure is accessible. They argue for the need to ethically compensate people for their data; accurately inform people about the estimated time and effort required to provide data (based on trials with people of the targeted group); and to be upfront about risks to privacy (also see [Ritter, 2021]). They also discuss the importance of collecting metadata that does not over simplify disability; and ensuring that disabled peoples' data is not unfairly rejected when minor mistakes occur or due to stringent time limits. Standard methods of data labeling, such as leveraging crowdworkers, have the potential to bake in biases about who is disabled, or what the meaning of disabled biometric data is. Whittaker (2019) discusses the example of clickworkers who label people as disabled "based on a hunch". Badly labeled data has many downstream implications for the quality, and potential negative impact, of biometric systems.

Park et al. (2021a) also advocate for the importance of *accessible data collection processes.* A basic requirement to improve data representation is making sure that data collection systems are accessible to everyone, and ensuring privacy and security of disability information in the specifics of how data is collected (*ibid.*). Similar expectations should be placed on each stage of data review, management, storage and monitoring.

Finally, it is important to ensure proper documentation. Abbott et al (2019) lay out guidelines for documentation and data security. Data management is a complex domain with many risks. While these concerns are universal, taking disability into account means ensuring that solutions to each of these challenges are accessible and open.

### d-f. Safeguards or limitations regarding approved use and mechanisms for preventing unapproved use; Performance auditing and post-deployment impact assessment; and Practices regarding the use of biometric technologies in conjunction with other surveillance technologies ( e.g., via record linkage);

At a basic level, just as websites are required to be accessible, so should algorithms. The W3C guidelines provide insight into website accessibility, but a similar set of expectations does not currently exist for biometrics. It is critical that we establish a basic set of expectations around how such algorithms are assessed for their accessibility. This should help to address **basic access constraints**, reduce the types of errors that **enforce "normality"** rather than honoring heterogeneity, and eliminate errors that gatekeep who is **"human"**.

Finally, as Ritter [2021] argues, consumer consent, and oversight around best practices, are both essential to fair use. Further, biometric systems should be interpretable and correctable, meaning that they can be overridden by a person based on their human judgment about a situation.  There should be particularly strong consequences when algorithms which are used to detect disability, or make decisions about access to services

9on the basis of disability, lack these properties. The potential consequences of errors made by these algorithms to health, safety, and participation in society are too severe to ignore.

## g-h. Practices or precedents for the admissibility in court of biometric information generated or augmented by AI systems and Practices for public transparency regarding: use, impacts, opportunities for contestation and for redress, as appropriate.

From how data is collected to how it is labeled to how it is used, it is critical that all stakeholders can participate in and understand their representation in biometric data. This requires that the data collection process be accessible, and that there is transparency about and documentation of what is collected and how it is used (Trewin, 2018). Transparency is critical to ensuring that all people can make safe and informed decisions about what services to use and when to take care or explore alternatives. It also incentivizes improvements in service quality. Further, transparency is critical to ensure that the rights of disabled people are enforceable in the court system (Whittaker, 2019, p. 17). Finally, as stakeholders with the same range of intelligence and commitment as anyone else, people considered and identifying themselves as disabled need to be in leadership positions. The slogan "Nothing about us without us" is not just memorable, but is how a just society works (Charlton, 1998).

## References

[ADA] U.S. Department of Justice, Civil Rights Division. The Americans with Disabilities Act (ADA).

[Abbott, 2019] Abbott, Jacob, et al. Local standards for anonymization practices in health, wellness, accessibility, and aging research at CHI. ACM *CHI* 2019.

[Araujo 2018] Araujo, M., et al. Ml approach for early detection of sleep apnea treatment abandonment: A case study. In the International Conference on Digital Health (pp. 75-79).

[Bennett 2020] Bennett, C., & Keyes, O. What is the point of fairness?. Interactions, 27(3), 35-39.

[Besmira, 2018] Nushi, B., et al. Towards accountable ai: Hybrid human-machine analyses for characterizing system failure. In *AAAI CHCC* 2018.

[Charlton, 1998] Charlton, James I. *Nothing about us without us*. University of California Press.

[DOJ 2014] United States Department of Justice, Statement of Interest in the New v. Lucky Brand Jeans (Apr. 10, 2014).

[Das 2012] Das, S., et al. Detecting Parkinsons' symptoms in uncontrolled home environments: A multiple instance learning approach. In IEEE Engineering in Medicine and Biology Society (pp. 3688-3691). IEEE.

[Demo, 2017] Demo, A. T. Hacking agency: Apps, autism, and neurodiversity. *Quarterly Journal of Speech*, *103*(3), 277-300.

[Guo, 2019] Guo, A., et al. Toward fairness in AI for people with disabilities: A research roadmap. ACM SIGACCESS Accessibility and Computing, (125), 1-1.

[Iwama, 2012] Iwama, Haruyuki, et al. The ou-isir gait database comprising the large population dataset and performance evaluation of gait recognition. In IEEE Transactions on Information Forensics and Security 7.5 (2012): 1511-1521.

[Kane, 2020] Kane, S. K., et al. Sense and accessibility: Understanding people with physical disabilities' experiences with sensing systems. In *ACM ASSETS*. 2020.

[Kasnitz, 2012] Kasnitz, D., & Block, P. (2012). Participation, time, effort and speech disability justice. Chapter 14, *Politics of occupation-centred practice: Reflections on occupational engagement across cultures*, Nick Pollard and Dikaios Sakellariou, Ed.John Wiley & Sons, Ltd.

[Larsen 2018] Loren Larsen, Keith Warnick, Lindsey Zuloaga, and Caleb Rottman, "Detecting Disability and Ensuring Fairness in Automated Scoring of Video Interviews," United States Patent Application Publication, August 20, 2018. Unearthed by AI Now Tech Fellow Genevieve Fried.

[Lazar, 2015] Lazar, J., et al. The discriminatory impact of digital inaccessibility. Chapter 3, *Ensuring Digital Accessibility Through Process and Policy.* Waltham, MA: Elsevier/Morgan Kaufmann Publishers.

[Leblanc 2020] Leblanc, E., et al. Feature replacement methods enable reliable home video analysis for machine learning detection of autism. Scientific reports, 10(1), 1-11.

[Lecher, 2018] Lecher, C. What happens when an algorithm cuts your health care, The Verge, 3/21/18

[Mittelstadt, 2019] Mittelstadt, B. Principles alone cannot guarantee ethical AI. Nature Machine Intelligence, 1(11), 501-507.

[Oliver, 2013] Mike Oliver. The social model of disability: Thirty years on. *Disability & society* 28(7):1024–1026.

[Park, 2021a] Park, J. S., et al. Designing an Online Infrastructure for Collecting AI Data From People With Disabilities. In the *ACM Conference on Fairness, Accountability, and Transparency* (pp. 52-63).

[Park, 2021b] Park, Joon Sung, et al. Understanding the Representation and Representativeness of Age in AI Data Sets. *arXiv preprint arXiv:2103.09058* (2021).

[Quintero, 2020] Quintero, Christian. A review: accessible technology through participatory design. *Disability and Rehabilitation: Assistive Technology* (2020): 1-7.

[Ritter, 2021] Ritter, E. Your Voice Gave You Away: the Privacy Risks of Voice-Inferred Information. Duke Law Journal, 71(3), 735-771.

[Taati, 2019] Taati, Babak, et al. Algorithmic bias in clinical populations—evaluating and improving facial analysis technology in older adults with dementia. IEEE Access 7 (2019): 25527-25534.

[Trewin 2018] Trewin, S. (2018). AI fairness for people with disabilities: Point of view. *arXiv preprint arXiv:1811.10670*. Also a blog post.

[Trewin, 2019] Trewin, S., et al. Considerations for AI fairness for people with disabilities. *AI Matters*, *5*(3), 40-63.

[UN 2007] UN General Assembly. (2007). Convention on the Rights of Persons with Disabilities: resolution / adopted by the General Assembly, 24 January 2007, A/RES/61/106. Retrieved Dec 2021.

[Whittaker 2019] Whittaker, M. et al. (2019). Disability, bias, and AI. *AI Now Institute*.

[Wolf, 2019] Wolf, Christine T., and Ringland, K. E. Designing accessible, explainable AI (XAI) experiences. *ACM SIGACCESS Accessibility and Computing* 125 (2019): 1-1.

[Wong, 2020] Wong, Alice, ed. *Disability visibility: First-person stories from the twenty-first century*. Vintage, 2020.

[Youngmann 2019] Youngmann, B., et al. A machine learning algorithm successfully screens for Parkinson's in web users. *Annals of clinical and translational neurology*, 6(12), 2503-2509.